\documentclass[letter]{jpsj2}

\newcommand{\kos}{KOs$_2$O$_6$}
\newcommand{\rbos}{RbOs$_2$O$_6$}
\newcommand{\csos}{CsOs$_2$O$_6$}
\newcommand{\cdre}{Cd$_2$Re$_2$O$_7$}
\newcommand{\cdos}{Cd$_2$Os$_2$O$_7$}

\newcommand{\msr}{$\mu$SR}

\title{Possible Anisotropic Order Parameter in Pyrochlore
Superconductor KOs$_2$O$_6$ Probed by Muon Spin Rotation}

\author{%
Akihiro \textsc{Koda}$^1$\thanks{E-mail address: coda@post.kek.jp}, %
Wataru \textsc{Higemoto}$^2$, %
Kazuki \textsc{Ohishi}$^1$, %
Shanta~R. \textsc{Saha},$^1$, %
Ryosuke \textsc{Kadono}$^1$\thanks{Also at %
School of Mathematical and Physical Science,
The Graduate University for Advanced Studies},
Shigeki \textsc{Yonezawa}$^3$, %
Yuji \textsc{Muraoka}$^3$ and Zenji \textsc{Hiroi}$^3$}
\inst{%
$^1$Muon Science Laboratory, Institute of Materials Structure
Science, High Energy Accelerator Research Organization (KEK),
1-1 Oho, Tsukuba, Ibaraki 305-0801\\
$^2$Advanced Science Research Center, Japan Atomic Energy Research
Institute, Tokai, Ibaraki 319-1195\\
$^3$Institute for Solid State Physics, University of Tokyo,
5-1-5 Kahiwanoha, Kashiwa, Chiba 277-8581
}

\abst{%
Magnetic penetration depth ($\lambda _{\rm eff}$) in a pyrochlore
 superconductor \kos\ was studied by means of muon spin rotation (\msr ).
The temperature dependence of the transverse spin relaxation rate, which is
 proportional to $\lambda _{\rm \rm eff}^{-2}$, exhibits a considerable deviation
 from that expected for the empirical two-fluid model ($\propto T^4$),
 obeying a $T^2$ tendency.
Furthermore, $\lambda _{\rm eff}$ apparently increases with increasing applied
 field, suggesting the presence of nonlinear and nonlocal effects.
The present result strongly suggests that
 the superconducting order parameter in \kos\ is anisotropic or
 multigapped with a small energy gap.
}

\kword{$\mu$SR, strongly correlated electron, anisotropic
superconductivity, pyrochlore oxide, flux-line lattice}

\begin{document}
\maketitle

Geometrically frustrated spin systems have been attracting much attention.
Since a three-dimensional network of corner-sharing tetrahedra, a so-called
pyrochlore lattice, exhibits magnetic frustration in the case of
antiferromagnetic exchange interaction between neighboring magnetic ions,
no magnetic ordering appears, leading to the {\em spin liquid} ground
state~\cite{Anderson56}.
It is noteworthy that electron transport under such strong magnetic
correlations exhibits interesting properties,
since the electron correlation also has an implicit relationship with the
charge and/or orbital degrees of freedom.
In particular, in the case of a mixed valence state, i.e., half-integer numbers
of $d$-electrons per magnetic site, it is suggested that charge/orbital
fluctuation plays an essential role.
For instance, the heavy-fermion-like behavior of LiV$_2$O$_4$ has been
reported~\cite{Kondo97}, in which each V ion forming a pyrochlore
lattice accommodates 1.5 $d$-electrons.
It has been shown that the electronic state exhibits strong correlation
under geometrical frustration~\cite{Koda04}.
This compound is one of only two spinel oxides showing metallic
conductivity without any structural transition even at low temperatures.
Another example is LiTi$_2$O$_4$ having 0.5 $d$-electrons per Ti ion,
which is known to be a superconductor with $T_c\simeq$~13~K~\cite{Johnston76}.
It has turned out, on the other hand, that many of the metallic compounds
having magnetic frustration show the metal--insulator (MI) transition upon
cooling.
As found in the cases of AlV$_2$O$_4$~\cite{Matsuno01}
and CuIr$_2$S$_4$~\cite{Radaelli02},
some compounds exhibit charge
ordering and an associated structural change that removes the geometrical degeneracy.
However, in other cases, e.g.,
Cd$_2$Os$_2$O$_7$~\cite{Sleight74,Mandrus01}, 
no structural transition is seen,
suggesting that the electronic properties are predominantly controlled by
magnetic frustration.

Recently, Yonezawa {\it et al.} discovered that \kos , which has a
$\beta$-pyrochlore structure, shows bulk superconductivity below
$T_c\simeq$~9.6~K~\cite{Yonezawa03}.
Moreover, a series of $\beta$-pyrochlore superconductors, \rbos\
($T_c\simeq$~6.3~K) and \csos\ ($T_c\simeq$~3.3~K), was subsequently
found~\cite{Yonezawa04a,Yonezawa04b}.
Until these discoveries, \cdre\ was the only pyrochlore oxide known to show
superconductivity ($T_c\simeq$~1~K)~\cite{Hanawa01,Sakai01,Jin01}.
As in the case of \cdre , a 5$d$ transition metal forms a pyrochlore
lattice in $\beta$-pyrochlore superconductors.
Nevertheless, it is notable that the formal oxidation state of the Os ion is
5.5+ (5$d^{2.5}$), which is specific to the $\beta$-pyrochlore structure,
while that of the Re ion in \cdre\ is 5+ (5$d^2$).
\cdos\ having Os$^{5+}$ (5$d^3$) exhibits the MI
transition at $\sim$230~K~\cite{Sleight74,Mandrus01}, suggesting the
presence of strong correlation among
$d$-electrons even in the case of \kos .

Furthermore, the geometrical frustration in \kos\ naturally gives
rise to the idea of superconductivity based on the {\em spin liquid}
state.
It is noteworthy that Na$_x$CoO$_2 \cdot y$H$_2$O, having a
two-dimensional triangular lattice of CoO$_6$, has been recently
identified as a superconductor with $T_c \simeq$~5~K~\cite{Takada03}.
The triangular lattice is well known as a stage of geometrical frustration.
Moreover, it is suggested from chemical titration and photoemission
studies that cobalt ions are in a mixed valence state
(Co$^{3.4+}$)~\cite{Sakurai}.

It is interesting to see whether or not the
superconducting state of \kos\ is similar to that of \cdre ,
because the resistivity in the normal state of \kos\ 
shows an anomalous broad shoulder with varying temperature, and no phase
transition down to $\sim$4~K has been confirmed by X-ray
diffractometry~\cite{Hiroi04},
which is in marked contrast to the case of \cdre ~\cite{Yamaura02}.
In this letter, we report the results of a muon spin rotation (\msr )
experiment on \kos\ under several transverse fields (TF) up to 6~T.
The observed temperature dependence of the magnetic penetration depth
$\lambda _{\rm eff}$ deduced from the TF-\msr\ measurement cannot be
explained by the empirical two-fluid model.
Furthermore, the field dependence of $\lambda _{\rm eff}$ strongly suggests
the existence of low-lying quasiparticle excitations, which are
attributed to either an anisotropic superconducting order parameter or a
multigap structure in \kos .

The \msr\ experiment was conducted at the M15 beamline of TRIUMF, Canada.
Time-dependent muon polarization under a transverse field was measured using a high-time-resolution \msr\
spectrometer furnished with a superconducting magnet.
For each \msr\ measurement, the magnetic field was applied at 20~K, which is well above $T_c$, and
subsequently the specimen was
cooled to the lowest temperature of 2~K.
The measured \kos\ sample was prepared as a pellet of polycrystalline
powder in a square shape of $8 \times 8$~mm$^2$.
Positive muons with a beam momentum of 29~MeV/c, whose polarizations were normal to the
beam axis, were implanted.
The $\mu ^+$-e$^+$ decay
asymmetry and associated Larmor precession with a frequency $\omega\equiv\gamma _{\mu}B$ (where
$\gamma _{\mu}=2\pi\times 135.54$~MHz/T is the muon gyromagnetic ratio) 
were measured to deduce the local field distribution.
The decay events in which muons missed the sample were eliminated with an
anticoincidence electronics logic circuit, resulting in quite low
background positron spectra.

It is well known that type-II superconductors show the flux-line lattice
(FLL) state at a field $B_{c1} < B < B_{c2}$, leading to the spatial
inhomogeneity of magnetic induction.
Since the implanted muons occupy random positions over the length
scale of FLL, i.e., $\lambda _{\rm eff}$, the observed precessing signal
$\hat{P}(t)$ is a spatial sum
of the internal field, given by
\begin{eqnarray}
\hat{P}(t) & \equiv & P_x(t)+ i P_y(t)=\int _{-\infty}^{\infty}
n(B) \exp (i\gamma _{\mu}Bt) dB,\\
n(B) & = & \left\langle\delta \left(B({\bf r})-B \right) 
\right\rangle _{\bf r},
\end{eqnarray}
where $n(B)$ is the spectral density for muon
precession determined by the local field distribution.
In polycrystalline samples, the Gaussian
distribution of local fields is a good approximation, namely
\begin{eqnarray}
\hat{P}(t) & \simeq & \exp (- \sigma ^2 t^2/2)\exp (i\gamma _{\mu}Bt),
\label{fitting}\\
\sigma & = & \gamma _{\mu} \sqrt{\langle\Delta B ^2\rangle},
\end{eqnarray}
where $\langle\Delta B ^2\rangle$ is the average of the second moment of
the field distribution ($\Delta B ^2 = [B(r) - B_0]^2$).
By simulating the distribution of $\lambda _{\rm eff}$, we confirmed that the above
assumption is reasonable, where the distribution is probably due to
randomness of FLL coming from various imperfections in a real
system~\cite{KadonoJPCM}.
The field variation for an ideal triangular FLL is
approximately estimated~\cite{Brandt88};
\begin{equation}
\sigma \left[\mu{\rm s}^{-1}\right]
\simeq 4.83 \times 10^{4} 
(1-b) \left[1+3.9(1-b)^2\right]^{\frac{1}{2}} 
\lambda _{\rm eff}^{-2} \left[{\rm nm}\right],
\label{lmd}
\end{equation}
where $b\equiv B/B_{c2}$.
Thus, $\lambda _{\rm eff}$ can be deduced directly from the measured relaxation rate $\sigma$~\cite{Sonier00}.

\begin{figure}[tb]
\begin{center}
\includegraphics[width=0.6\textwidth]{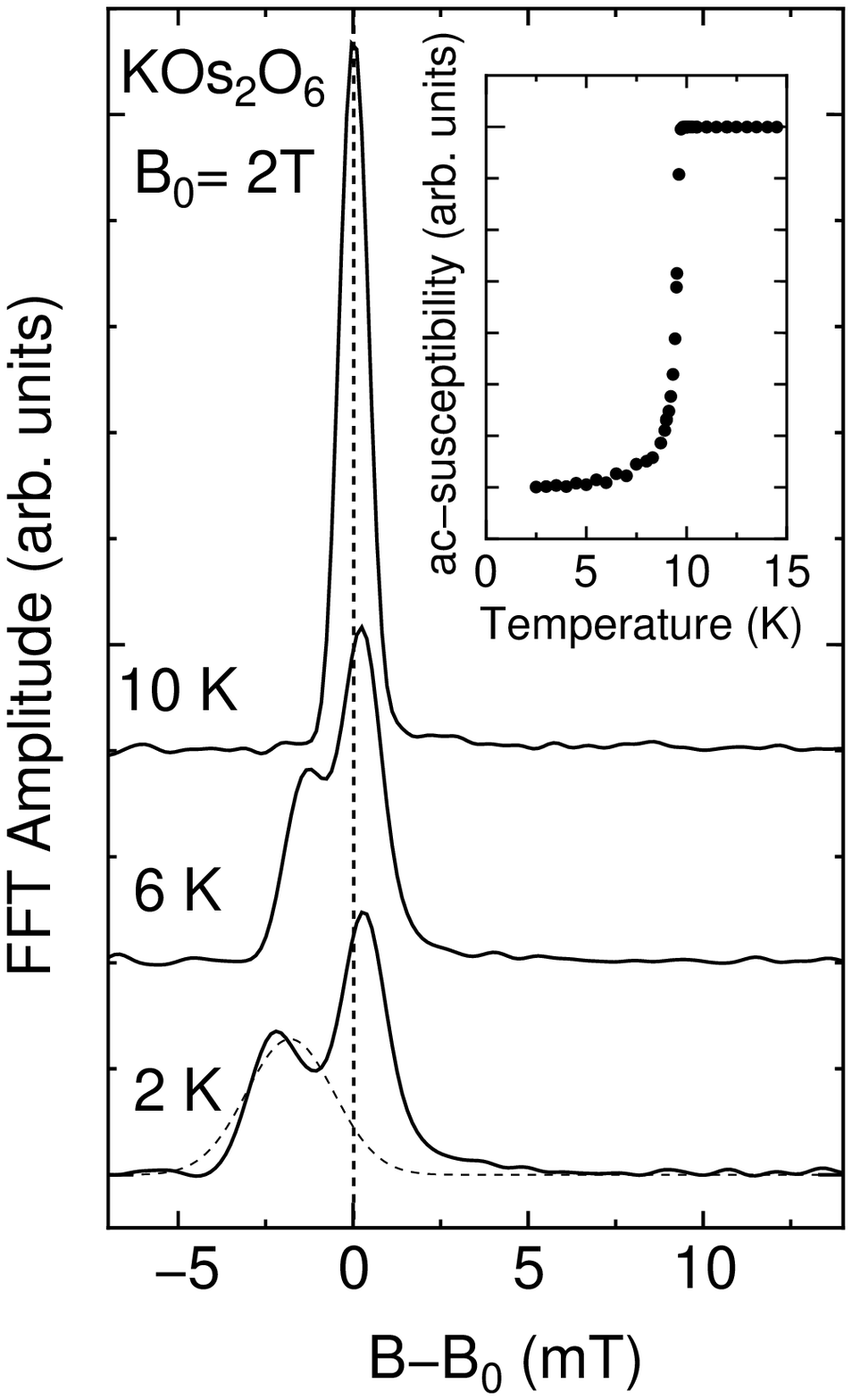}
\end{center}
\caption{%
Fast Fourier transform (FFT) of $\mu$SR time spectra under field
 of 2~T at various temperatures.
The dashed curve represents the deduced field distribution of the
 superconducting part obtained by the fitting analysis at $T=$2~K.
Inset shows the ac-susceptibility of the employed specimen.}
\label{spectra}
\end{figure}

\begin{figure}[tb]
\begin{center}
\includegraphics[width=\textwidth]{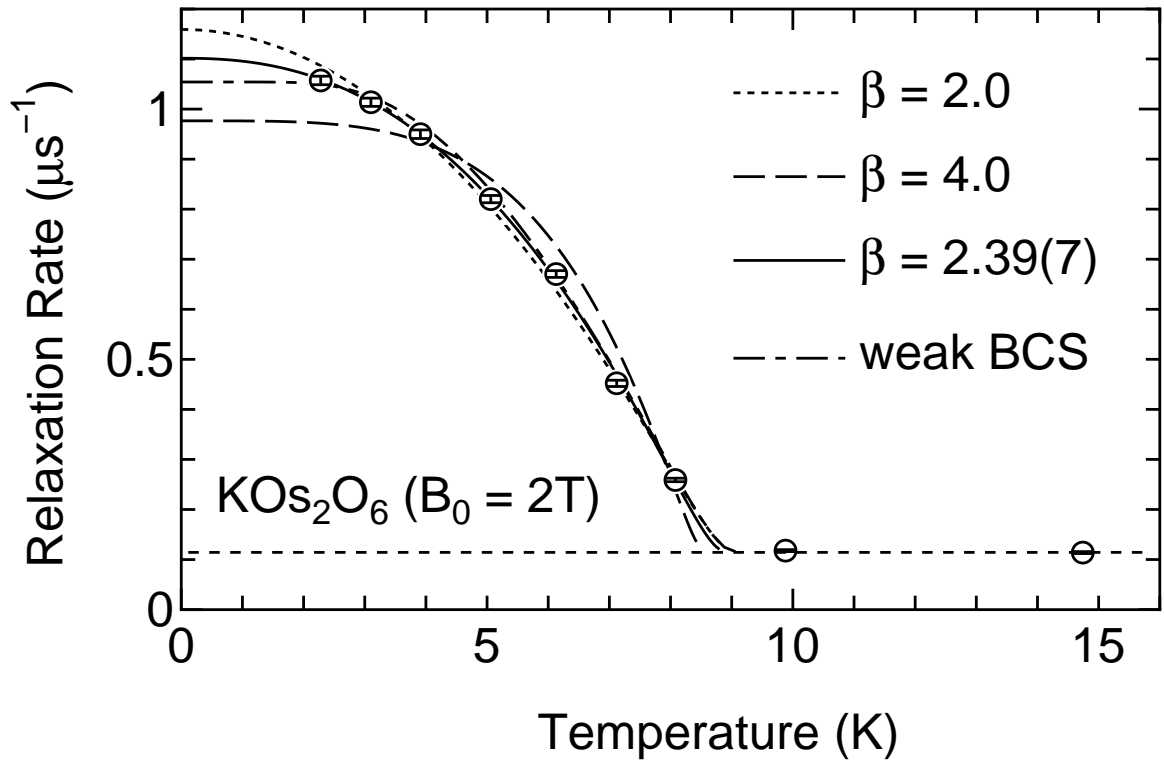}
\end{center}
\caption{%
Temperature dependence of transverse spin relaxation rate ($\sigma$) under
field of 2~T.
The results of fitting analysis using the relation
 $\sigma\propto 1-(T/T_c)^{\beta}$ are shown, together with the
 weak-coupling BCS case (dot-dashed curve).
The dotted curve is for $\beta = 2$, whereas the dashed curve is 
for $\beta = 4$.
The best fitting is obtained when $\beta = 2.39$, which is represented
 by the solid curve.
}
\label{t-sweep}
\end{figure}

In Fig.~\ref{spectra}, we show the fast Fourier transform (FFT) of the
obtained TF-\msr\ time spectra at different temperatures.
The line shape exhibits apparent broadening and a shift of the peak to lower
frequencies below $T_c \simeq$~9~K, which is ascribed to the formation of the FLL
state.
On the other hand, there exists a relatively narrow peak at around the
central frequency corresponding to the applied field, clearly
suggesting that a nonsuperconducting (normal) portion was involved in the measured sample.
The peak slightly moves to higher frequency below $T_c$.
Such behavior is often seen in the case of polycrystalline
superconductors, e.g., MgB$_2$~\cite{Ohishi03}, due to the local
demagnetization of grains.
The volume fraction of the normal part is estimated by fitting analysis
to be about 20\%, which is in line with that found by the bulk
susceptibility measurement~\cite{Yonezawa03}.
After the \msr\ measurement, it turned out that the employed sample
involves a moderate amount of KOsO$_4$, which is an insulator.
The analysis of the \msr\ experiment was carried out on the time-differential
data using the sum of the two precessing parts given by
eq.~\ref{fitting}.
Thus, we can obtain the field distribution solely ascribed to the
superconductivity of \kos .
The temperature dependence of the muon spin relaxation rate of the
superconducting part is shown in Fig.~\ref{t-sweep}.
In general, the London penetration depth $\lambda _L$ is given by the following relation
\begin{equation}
\lambda _L^2 = \frac{m^{\ast}c^2}{4\pi n_s e^2},
\label{l-general}
\end{equation}
where $m^{\ast}$ is the effective mass and $n_s$ is the superconducting
carrier density.
According to the Gorter-Casimir two-fluid model,
the superconducting carrier density $n_s$ is proportional to
$1-(T/T_c)^4$.
Thus, we have $\lambda _L\propto 1/\sqrt{1-(T/T_c)^4}$,
leading to $\sigma\propto 1-(T/T_c)^4$.
As shown in Fig.~\ref{t-sweep}, our data exhibit significant deviation
from this relation.
The fitting analysis was then performed with an arbitrary power
\begin{equation}
\sigma = \sigma _{T=0} \left[1- \left(\frac{T}{T_c}\right)^{\beta}\right],
\end{equation}
with $T_c$ as a free parameter.
Here, the contribution from the nuclear magnetic moments such as $^{39}$K
and $^{189}$Os seen at $T>T_c$
was subtracted by a fixed amount prior to the fitting analysis.
While the best fitting is obtained when $\beta = 2.39(7)$ and 
$T_c = 8.91(3)$, the observed temperature dependence is nearly
reproduced by $\beta = 2$ and $T_c = 9.07(2)$.
The results are shown in Fig.~\ref{t-sweep} together with the fitting
result by the weak-coupling BCS model~\cite{Muhlschlegel59}.
Note that $T_c$ deduced from the \msr\ measurement is fairly close
to that determined by bulk measurements under a field of
2~T~\cite{Hiroi04}.
Besides, as seen in Fig.~\ref{t-sweep}, we have to note that
distinguishing such power-law behavior, which may indicate the presence
of low-lying excited states, from the fully gapped case (weak
BCS) only by analyzing the present data points above 2~K is quite difficult.

\begin{figure}
\begin{center}
\includegraphics[width=0.8\textwidth]{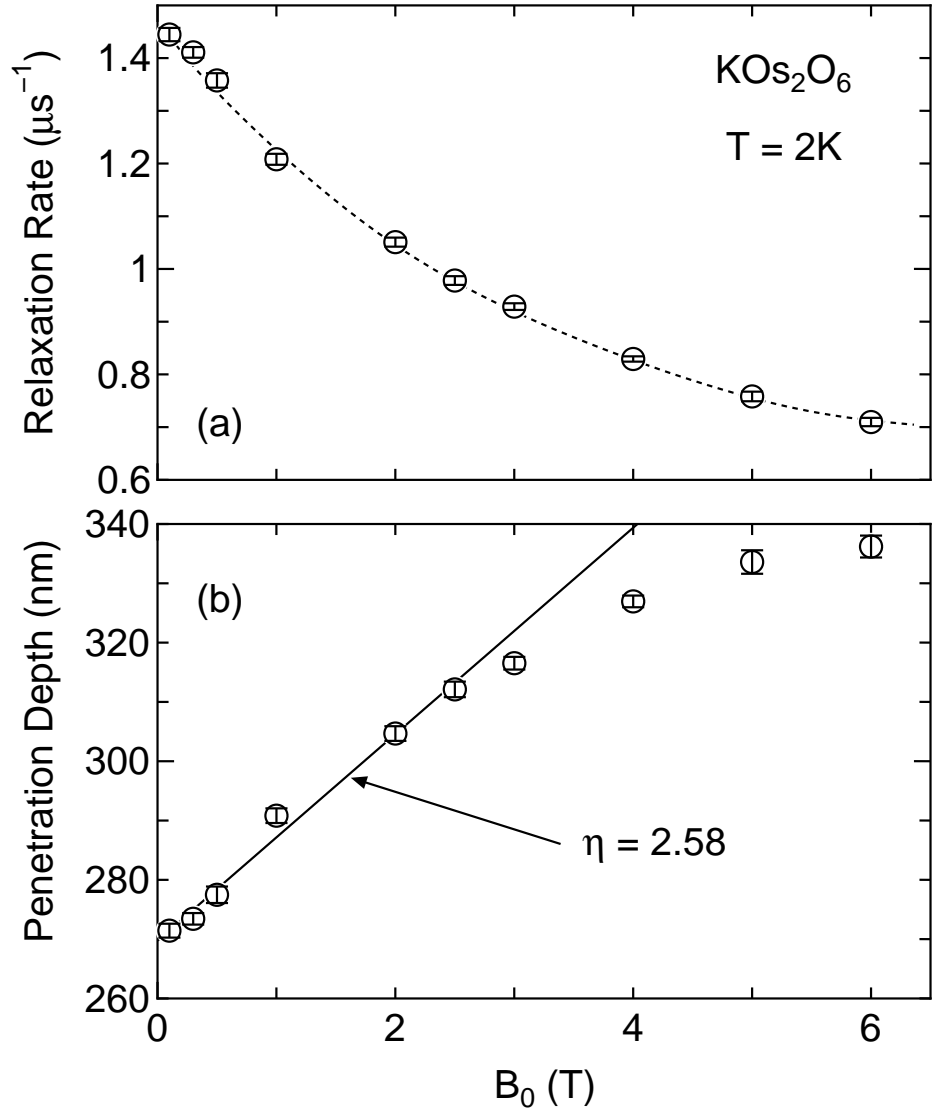}
\end{center}
\caption{%
Field dependences of (a) the transverse relaxation rate at 2~K and (b)
 the deduced penetration depth ($\lambda _{\rm eff}$) by assuming
 $B_{c2}=$ 40~T.
The broken curve in the upper panel is a guide to the eye.
$\lambda _{\rm eff}$ clearly obeys a linear dependence below $B \simeq$ 3~T,
 whereas it shows a saturating tendency at higher fields.
By fitting data below 2.5~T with the relation
$\lambda _{\rm eff} = \lambda _0 (1+ \eta (B/B_{c2}))$,
we obtain $\eta =$ 2.58.}
\label{h-sweep}
\end{figure}

In order to examine the low-lying quasiparticle excitations of \kos ,
we measured the field dependence of $\lambda _{\rm eff}$ at the lowest
temperature of 2~K.
In Fig.~\ref{h-sweep}(a), transverse relaxation rates at various fields
are shown.
Although $\lambda _{\rm eff}$ can be evaluated using eq.~\ref{lmd}, the exact
value of $B_{c2}$ in the examined specimen is still unknown.
Assuming that $B_{c2}\simeq$ 40~T~\cite{Hiroi04}, we deduced
$\lambda _{\rm eff}$ versus the corresponding normalized field, as shown in Fig.~\ref{h-sweep}(b).
Apparently, the obtained $\lambda _{\rm eff}$ is field dependent, showing
a slight negative curvature with respect to the applied field.
In particular, an approximately linear tendency is seen below $\sim$3~T.

In general, $\lambda _L$ is enhanced under a sufficiently high field,
because the superconducting carrier density $n_s$ decreases due to
pair-breaking
interactions such as the Zeeman interaction.
However, the magnitude of this effect is small in conventional
superconductors with {\em isotropic} $s$-wave pairing.
On the other hand, a strong enhancement of $\lambda _{\rm eff}$ with external field
is expected in the case of superconductors with the anisotropic order
parameter due to the combined
contribution of (i) nonlinear and (ii) nonlocal
effects.
More specifically, supercurrent ${\bf v}_s$ induced by the magnetic field causes a
semiclassical Doppler shift of the quasiparticle energy levels, which is
proportional to ${\bf v}_s\cdot {\bf v}_F$, where ${\bf v}_F$ is the
Fermi velocity. 
This gives rise to a nonlinear response of the shielding current
to the field due to the ``backflow''~\cite{Bardeen54}.
More importantly, the Doppler shift enhances the quasiparticle
excitation in the case of anisotropic pairing symmetry due to the
low-lying quasiparticle excitations.
Note that the excess quasiparticles due to the thermal activation at a
finite temperature exist in the vicinity of gap minimum even in the case
of {\em anisotropic} $s$-wave pairing.
No such enhancement is expected
in the case of {\em isotropic} $s$-wave superconductors as long as the
gap energy is greater than the energy of the Doppler-shifted
quasiparticles.
Thus, the present result clearly exhibits the existence of low-lying
quasiparticle excitations associating with the superconducting energy
gap smaller than 2~K ($\simeq$0.2~meV).

When the superconducting energy gap has nodes at
the Fermi surface, the field dependence of $\lambda _{\rm eff}$ is
further modified by the nonlocal effect.
The coherence length $\xi _k$ diverges at the nodes, because $\xi _k$ is
inversely proportional to the gap size
$\xi _k = \left|{\bf v}_F \right| / \pi \Delta _k$,
where $\Delta _k$ is the angular-dependent energy gap at the Fermi
surface, resulting in the reformation of the
FLL.
As shown in Fig.~\ref{h-sweep}(b), $\lambda _{\rm eff}$ tends to exhibit
saturating behavior at higher fields.
It is interesting to point out that a similar tendency has been discussed in the
case of YBa$_2$Cu$_3$O$_{7-\delta}$ by considering nonlinear and
nonlocal effects, both theoretically~\cite{Amin98,Amin00} and
experimentally~\cite{Sonier00,Sonier99}.
In any case, we would like to emphasize that the field-dependent behavior of
$\lambda _{\rm eff}$ is characterized by a small energy scale less than 0.2~meV.

To evaluate the strength of the pair-breaking effect, we performed a fitting using
a simple linear relation
\begin{equation}
\lambda _{\rm eff} = \lambda _0 [1+ \eta b].
\end{equation}
From the analysis of data below 3~T, we obtain $\lambda _0 =$ 270~nm and
$\eta =$ 2.58 ($T/T_c \simeq 0.22$), which is shown as the solid
line in Fig.~\ref{h-sweep}(b).
We stress that $\lambda _{\rm eff}$ is independent of the applied field in the case
of {\em isotropic} $s$-wave pairing symmetry.
In V$_3$Si, which is a typical $s$-wave superconductor,
$\eta\simeq 0$ is confirmed at lower fields ($B/B_{c2}<0.5$)~\cite{Sonier04}.
This is in marked contrast to the present case.
%
Besides this, it has turned out that MgB$_2$ exhibits $\eta =1.27$ at
0.26$T_c$~\cite{Ohishi03}.
In this case, it is suggested that the presence of two superconducting
energy gaps, one being relatively small ($\sim$1~meV), is responsible for the
field dependence of $\lambda _{\rm eff}$ observed at a finite temperature.
In this sense, we cannot exclude the multiple-gap scenario in \kos , since the
measurement was performed at a finite temperature of 2~K.
The present result
suggests the presence of a small energy gap having
a magnitude of less than $\sim$0.2~meV in the multiple-gap scenario.
It is noted, however, that distinguishing between the anistropic-gap case
and the multiple-gap case is difficult with our present experimental data.
In order to clarify this, a measurement at a further low temperature is planned.
On the other hand, the finite value of $\eta = 2.58$ in \kos\ is
obviously different from that reported in the case of \cdre
~\cite{Kadono02,Lumsden02}($\eta\simeq 0$) in spite of the structural
similarity.
This suggests that structural
transitions found in \cdre\ ~\cite{Yamaura02} have a crucial effect on
the pairing mechanism.
In contrast, the absence of structural transitions in \kos\ suggests
that the geometrical degeneracy remains even at low temperatures,
leading to a {\em spin liquid} state, in which a dynamical magnetic
correlation exists.
One may expect that the Cooper pair is mediated by magnetic fluctuation
under such circumstances.


In summary, we have investigated the quasiparticle excitation in \kos\ by
measuring the magnetic penetration depth $\lambda _{\rm eff}$.
The temperature dependence of $\lambda _{\rm eff}$ exhibits a significant
deviation from that expected for the empirical two-fluid model.
Moreover, it was seen that $\lambda _{\rm eff}$
increases markedly with applied external field up to 6~T, suggesting the
presence of nonlinear and nonlocal effects.
We consider this to be evidence that the
superconducting order parameter realized in \kos\ has a strong
anisotropy or multigapped structure with a small gap energy.

We are grateful for the technical support of the TRIUMF staff during the
experiment.
This work was partially supported by a Grant-in-Aid for Scientific
Research on Priority Areas and a Grant-in-Aid for Creative Scientific
Research from the Ministry of Education, Culture, Sports,
Science and Technology, Japan.
One of the authors (A.K.) acknowledges the support of JSPS Research Fellowships for Young Scientists.


\end{document}